\documentclass[a4paper,12pt]{article}

\usepackage{epsfig}

\usepackage{amsmath}
\usepackage{amssymb}
\usepackage{cite}

\usepackage{fancyhdr}
\newcommand{\beqs}{\begin{equation*}}
\newcommand{\beq}{\begin{equation}}

\newcommand{\eeqs}{\end{equation*}}
\newcommand{\eeq}{\end{equation}}

\newcommand{\beqas}{\begin{eqnarray*}}
\newcommand{\beqa}{\begin{eqnarray}}

\newcommand{\eeqas}{\end{eqnarray*}}
\newcommand{\eeqa}{\end{eqnarray}}

\newcommand{\eq}[2]{\begin{equation} #1 \label{#2} \end{equation}}

\newcommand{\meq}[2]{\begin{multline} #1 \label{#2} \end{multline}}

\newcommand{\eps}{\varepsilon}
\newcommand{\al}{\alpha}

\newcommand{\ga}{\gamma}
\newcommand{\de}{\delta}
\newcommand{\om}{\omega}

\newcommand{\Om}{\Omega}

\newcommand{\blist}{\begin{itemize}}

\newcommand{\elist}{\end{itemize}}

\providecommand{\href}[2]{#2}

\DeclareFontFamily{OT1}{rsfs}{}
\DeclareFontShape{OT1}{rsfs}{m}{n}{ <-7> rsfs5 <7-10> rsfs7 <10->rsfs10}{} 
\DeclareMathAlphabet{\mycal}{OT1}{rsfs}{m}{n}

\providecommand{\href}[2]{#2}

\newcommand{\mom}{{\mathcal T}}

\begin{document}

\begin{titlepage}

\renewcommand{\thefootnote}{\fnsymbol{footnote}}

\hfill TUW--03--26

\hfill ESI 1423


\begin{center}
\vspace{0.5cm}

{\Large\bf The ultrarelativistic limit of 2D dilaton gravity and its energy momentum tensor}
\vspace{1.0cm}

{\bf H.\ Balasin\footnotemark[1], D.\ Grumiller\footnotemark[2]\footnotemark[3]}
\vspace{7ex}

  {\footnotemark[1] \footnotemark[2]
\footnotesize Institut f\"ur
    Theoretische Physik, Technische Universit\"at Wien \\ Wiedner
    Hauptstr.  8--10, A-1040 Wien, Austria}
\vspace{2ex}

  {\footnotemark[3]\footnotesize Institut f\"{u}r Theoretische Physik, Universit\"{a}t Leipzig,\\Augustusplatz 10, D-04109 Leipzig, Germany}

   \footnotetext[1]{e-mail: \texttt{hbalasin@tph.tuwien.ac.at}}
   \footnotetext[2]{e-mail: \texttt{grumil@hep.itp.tuwien.ac.at}}
\end{center}
\vspace{7ex}

\begin{abstract}


The ultrarelativistic limit of twodimensional dilaton gravity is presented and its associated (anti-)selfdual energy momentum tensor is derived. It is localized on a null line, although the line element remains twice differentiable. Relations to the Aichelburg-Sexl spacetime and constant dilaton vacua are pointed out. Geodesics are found to be smooth for minimally coupled test particles but non-smooth -- with a finite jump in the acceleration -- for test particles coupled non-minimally to the dilaton. Quantization on boosted backgrounds is discussed; no anomalous trace of the energy momentum tensor arises and the 1-loop flux component can be adjusted to be equal to the classical flux of the shock wave. 

\end{abstract}


\vfill
\end{titlepage}

\section{Introduction}
In their by now classical work \cite{Aichelburg:1971dh} Aichelburg
and Sexl (AS) calculated the ultrarelativistic (UR) limit of the Schwarzschild
geometry. That is to say the form of the gravitational field as it
appears to an asymptotic observer moving very rapidly relative to
the the black hole. It turns out that the gravitational field is completely
concentrated on a null-hyperplane and that the spacetime is flat
above and below the pulse. More precisely the geometry represents
an impulsive pp-wave \cite{Jordan:1960}
\eq{
ds^{2}=2dudv-d\tilde{x}^{2}+8p\delta(u)\log(\rho)du^{2}\,,\quad\rho:=\sqrt{\tilde{x}^i\tilde{x}_i}\,\,,
}{eq:AS}
where $u,v$ denote linear null coordinates and $\tilde{x}$ refers to the transversal spacelike coordinates in Minkowski space. The mass parameter $p$ defines the physical scale. The
matter content is localized on a single generator of the
pulse-plane, which immediately gives rise to the dual interpretation
of the AS-geometry as the gravitational field of massless particle
in Minkowski space. The latter interpretation inspired 'tHooft and
Dray \cite{Dray:1984ha} to a remarkable generalization of the AS-geometry.
They were able to construct a solution of the Einstein equations 
describing the change in the gravitational field of a Schwarzschild
black hole due to a massless particle that moves along a particular generator of
the horizon.

Both interpretations have inspired various generalizations like UR
limits of other black hole geometries \cite{Lousto:1988ua,Balasin:1995tb,Balasin:1995tj} 
as well as changes in gravitational fields due to a massless particle
\cite{Alonso:1987} 
where general conditions for the
applicability of the 'tHooft-Dray procedure where clarified in ref.\ \cite{Balasin:1999wg}.
In the present work we will use the second interpretation to construct
impulsive solutions of twodimensional dilaton gravity. (In the appendix
we will briefly comment on the first interpretation)

This work is organized as follows: sect.\ \ref{se:2} briefly reviews dilaton gravity in two dimensions. In sect.\ \ref{se:3} the UR limit is treated: the energy-momentum 1-form is derived by the patching procedure and the behaviour of geodesics of test-particles is discussed. Sect.\ \ref{se:qu} is devoted to quantization of and on such geometries. The final sect.\ \ref{se:4} contains a brief summary and possible generalizations of the obtained results.

\section{Classical dilaton gravity}\label{se:2} 

In the 1990-ies the interest in dilaton gravity in two dimensions ($2D$) was rekindled by results from string theory \cite{Mandal:1991tz,Callan:1992rs}, 
but it existed as a field on its own more or less since the 1980-ies \cite{Barbashov:1979}. 
There are (at least) four different motivations to study dilaton gravity in $2D$: 1.~The s-wave part of General Relativity can be reproduced and therefore be treated in a more transparent form, 2.~Certain dilaton gravity models appear in the context of string theory (most recently in $2D$ type 0A/0B; cf.~the somewhat randomly collected list of references \cite{Douglas:2003up}), 
3.~Technically speaking, solubility increases in lower dimensional models which in turn allows the discussion of non-perturbative aspects, and thus certain toy models serve as a convenient laboratory for quantizing gravity and for studying black hole evaporation, 4.~From a purely mathematical point of view, in a first order formulation the underlying Poisson structure reveals relations to non-commutative geometry and deformation quantization. In this sense, dilaton gravity even may serve as a link between General Relativity, string theory, black hole physics and non-commutative geometry.
For a more detailed discussion and further historical details the review ref.\ \cite{Grumiller:2002nm} may be consulted. For sake of self-containment the study of dilaton gravity will be motivated briefly from a purely geometrical point of view.

The notation of ref.\ \cite{Grumiller:2002nm} is used: $e^a=e^a_\mu dx^\mu$ is the
dyad one-form dual to $E_a$ -- i.e.\ $e^a(E_b)=\de^a_b$. Latin indices refer to an anholonomic frame, Greek indices to a holonomic one. The one-form
$\omega$ represents the  spin-connection $\om^a{}_b=\eps^a{}_b\om$
with  the totally antisymmetric Levi-Civit{\'a} symbol $\eps_{ab}$ ($\eps_{01}=+1$). With the
flat metric $\eta_{ab}$ in light-cone coordinates
($\eta_{+-}=1=\eta_{-+}$, $\eta_{++}=0=\eta_{--}$) the torison 2-form reads
$T^\pm=(d\pm\omega)\wedge e^\pm$. The curvature 2-form $R^a{}_b$ can be represented by the 2-form $R$ defined by 
$R^a{}_b=\eps^a{}_b R$, $R=d\wedge\om$. The volume two-form is denoted by $\epsilon = e^+\wedge e^-$. Signs and factors of the Hodge-$\ast$ operation are defined by $\ast\epsilon=1$. 

Since the Einstein-Hilbert action $\int_{\mathcal{M}_2}  R\propto(1-g)$ yields just the Euler number for a surface with genus $g$ one has to generalize it appropriately. The simplest idea is to introduce a Lagrange multiplier for curvature, $X$, also known as ``dilaton field'', and an arbitrary potential thereof, $V(X)$, in the action $\int_{\mathcal{M}_2}  \left(XR+\epsilon V(X)\right)$. In particular, for $V\propto X$ the Jackiw-Teitelboim model emerges \cite{Barbashov:1979}. 
Having introduced curvature it is natural to consider torsion as well. By analogy the first order gravity action \cite{Schaller:1994es}
\eq{
\int_{\mathcal{M}_2}  \left(X_aT^a+XR+\epsilon\mathcal{V} (X^aX_a,X)\right)
}{eq:FOG}
can be motivated where $X_a$ are the Lagrange multipliers for torsion. It encompasses essentially all known dilaton theories in $2D$.

For simplicity the discussion will be restricted to the so-called $a-b$ family \cite{Katanaev:1997ni} of dilaton gravity models, i.e.\ actions of the type
\eq{
L=\int_{\mathcal{M}_2} \left[X_a (d\pm\om)\wedge e^a +Xd\wedge\omega + \epsilon \left( U(X)X^+X^-+V(X)\right) \right]\,,
}{cs:1}
with 
\eq{
U(X)=-\frac{a}{X}\,, \quad V(X)=-\frac{B}{2}X^{a+b}\,,\quad a,b,B\in\mathbb{R}\,.
}{cs:1.5}
Among a variety of other interesting models also the spherically reduced Schwarz\-schild black hole (BH) belongs to this class ($a=1/2=-b$). There can be at most one (nonextremal) Killing horizon for each solution of such a model. We will comment on the generic case in sect.\ \ref{se:4}.

In the absence of matter all classical solutions can be constructed locally and globally \cite{Klosch:1996fi}. 
In a patch where $X^+\neq 0$ the solution can be presented in generalized Kerr-Schild form:
\eq{
(ds)^2=g_{\mu\nu}(r)dx^\mu dx^\nu - f(r)(du)^2\,, \quad f=2MX^{-a}\,,\quad dr=X^{-a}dX
}{boost:2}
Evidently there is always a Killing vector $\xi^\al\partial_\al=\partial/\partial u$. The constant $M$ is proportional to the Casimir function in the language of first order gravity and to the ADM mass in the language of General Relativity (whenever this concept is well-defined, i.e.\ asymptotic flatness). The light-like coordinate $u$ has been chosen in outgoing form for sake of definiteness. The radial coordinate $r$ and the dilaton $X$ are related by a coordinate transformation which is singular at the origin and in the asymptotic region. The background metric of the ground state geometry ($M=0$) in outgoing Eddington-Finkelstein gauge is given by
\eq{
g_{\mu\nu}(r)dx^\mu dx^\nu = 2dudr + \frac{B}{(b+1)}X^{b-a+1} (du)^2\,,\quad b\neq -1\,.
}{boost:3}
For Minkowski Ground State (MGS) models the relation $a=b+1$ guarantees a flat background metric and thus reduces from generalized Kerr-Schild type to ordinary one. The curvature scalar associated with the Levi-Civit\'a connection (but in general, i.e.\ for $a\neq 0$, it is {\em not} associated with the connection $\om$)
\eq{
R=-\frac{d^2 f}{dr^2}+\frac{Bb}{(b+1)}(b-a+1)X^{a+b-1}
}{boost:4}
has potential singularities only at $X=0$ and/or at $X=\infty$, depending on the specific values of $a,b$ and (to a lesser extent) $B$. In the derivation of (\ref{boost:3}) it has been assumed that $X^+\neq 0$. It is not possible to impose the same requirement on $X^-$ because it changes sign at a Killing horizon. 

\section{Ultrarelativistic dilaton gravity}\label{se:3}

Because the equations of motion (\ref{eq:a5})-(\ref{eq:a9}) are obviously invariant under boosts 
one has to be careful in constructing a meaningful UR limit. In fact, one has to {\em define} properly what is meant by ``UR limit''. 

Suppose the considered geometry exhibits a Killing horizon. Then, also all boosted geometries will show this feature. However, eventually we would like to peform an infinite boost and at the same time let $M\to 0$ such that a nontrivial limit emerges.\footnote{Clearly, letting just $M\to 0$ yields something trivial. On the other hand, taking the UR limit without scaling $M$ to zero yields a singular spacetime.} It should be emphasized that boosting the spherically reduced Schwarzschild BH is quite different from boosting its fourdimensional ancestor, because the latter, when boosted, explicitly breaks spherical symmetry, while the former by construction always remains spherically symmetric (the employed boosts are radial ones). Therefore, methods which work well in four dimensions need not be suitable in the twodimensional realm.

A possible strightforward definition is to consider boosts with respect to the background metric (\ref{boost:3}), $du\to du\exp{\ga}$, $dv\to dv\exp{(-\ga)}$ with $dv=dr+du\cdot X^{b-a+1}B/(2(b+1))$, to take the limit $\ga\to\infty$, and to fix the scaling behavior of $M$ in (\ref{boost:2}) such that the whole term $f(r)(du)^2$ survives the limit in a well-defined way \cite{Aichelburg:1971dh}. Unfortunately, this route is rather difficult to pursue for generic dilaton gravity, because, for instance, solutions need not be asymptotically flat. 

Alternatively, one can try to obtain the curvature scalar as a well-defined distribution and use it to construct a distributional energy-momentum tensor. Then, the UR limit can be obtained relatively easy \cite{Balasin:1995tb}. Although the first part is possible to a certain extent (cf.\ appendix \ref{app:B}) the second part fails because curvature is not related directly to the energy-momentum tensor (cf.\ appendix \ref{app:A}). Instead, the energy-momentum 1-form is generated by derivative terms of the auxiliary fields $X^\pm$ in eq.\ (\ref{eq:a6}). Thus, it seems to be a good idea to construct the UR limit from this perspective, i.e.\ to focus on the geometric properties\footnote{If the twodimensional theory has been obtained by reducing a higherdimensional Einstein theory then the auxiliary fields $X^\pm$ are typically certain components of the higherdimensional spin connection. For instance, for spherically reduced gravity they enter the 1-form $\om^i{}_a \propto (E_a X)e^i$, where $a=\pm$ and $i=\theta,\phi$ (cf.\ e.g.\ appendix A of ref.\ \cite{Grumiller:2002nm}). For many practical purposes one can think of $X^\pm$ as the expansion spin coefficients $\rho$ and $\rho'$ (both are real).} of $X^\pm$.

\subsection{The patching procedure}

\begin{figure}
\centering
\epsfig{file=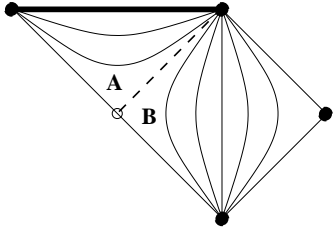,width=0.32\linewidth}
\caption{Single EF patch}
\label{fig:EF}
\end{figure}

\begin{figure}
\centering
\epsfig{file=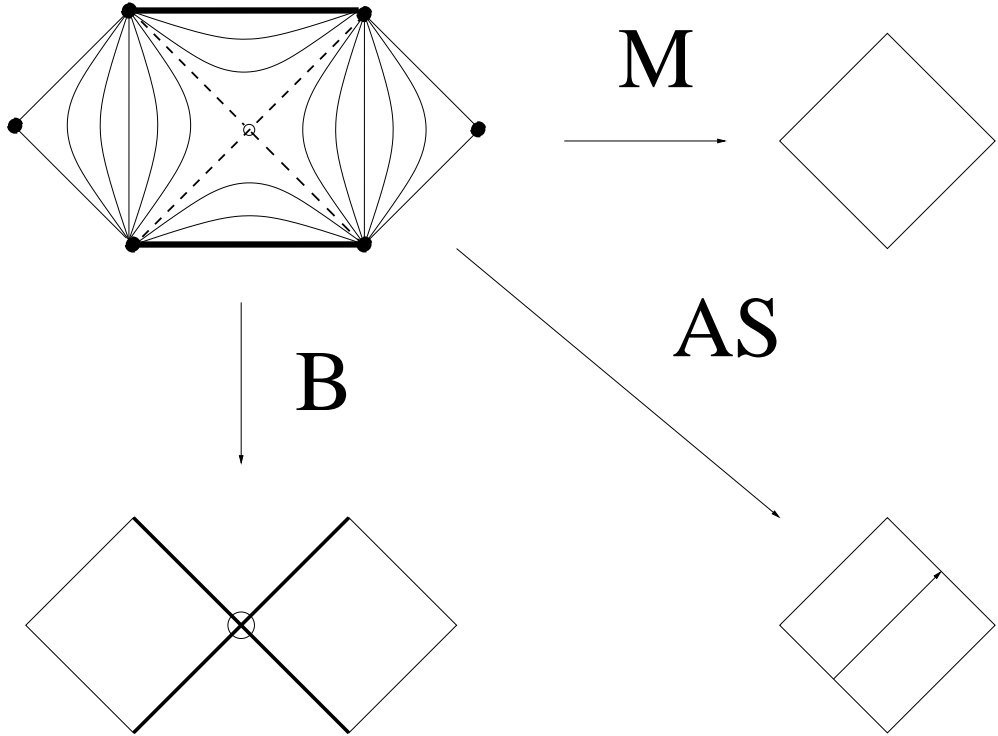,width=0.96\linewidth}
\caption{Various limits of the Schwarzschild BH: $B$ means infinite boost (yielding a rather pathological spacetime where the singularity coincides with the horizon), $M$ implies mass to zero (providing Minkowski space as limit), and $AS$ is the AS limit where simultaneously the boost parameter tends to infinity and the mass to zero (again one obtains Minkowski space except for a certain value of the outgoing EF coordinate $u=u_0$ where a shockwave propagates)}
\label{fig:limits}
\end{figure}

The following observation is very helpful in constructing the UR limit: the global diagram of the Schwarzschild BH can be interpreted as a double covering of a single Eddington-Finkelstein (EF) patch (fig.\ \ref{fig:EF} or a mirror version thereof; one of the Lagrange multipliers for torsion has a definite sign in the whole patch while the other one changes its sign at the Killing horizon -- for instance, in region $B$ ($A$) $X^->0$ ($X^-<0$), while $X^+>0$ in the whole patch).
In the AS limit the mass goes to zero, the horizon vanishes, but the double covering structure is kept {\em per definitionem}. The possible limiting geometries are displayed in fig.\ \ref{fig:limits}.

The distinguishing features of the EF patches in the context of dilaton gravity are the signs of $X^\pm$: in any one of the four possible patches (each of which covering half of the spacetime) either $X^+$ or $X^-$ has a definite sign, while the other quantity changes its sign on the Killing horizon. Thus, we propose to construct the UR limit as follows: take a classical solution of dilaton gravity in a patch with, say, $X^+>0$; take the limit $M\to 0$; repeat the same with a patch where $X^+<0$; join these two solutions and calculate the induced matter flux along the matching surface where $X^+=0$. 

\subsection{The energy-momentum tensor}

The crucial observation is that such a patching implies discontinuity of the quantity $X^+$,
\eq{
X^+=\frac{X^+_0}{2} \left(\theta(u-u_0)-\theta(u_0-u)\right)\,,\quad X^+_0\in\mathbb{R}^+\,.
}{boost:6}
Eq.\ (\ref{eq:a6}) then induces the existence of a localized energy-momentum 1-form\footnote{If one supposes the induced matter to be a Klein-Gordon field then $\mom^-=0$ imposes anti-selfduality. The localization of the energy-momentum tensor implies that the field itself does not produce a meaningful limit as a distribution (it would be something like the integral over the square-root of the $\de$-function). However, its energy-momentum tensor obviously is a well-defined distribution.}
\eq{
\mom^+=-X^+_0\de(u-u_0)du\,,\quad \mom^-=0\,.
}{boost:7}
This resembles the analogue expression in four dimensions \cite{Balasin:1995tb} if presented as
\eq{
\mom^a{}_b\propto \de(p\cdot x)p^ap_b\,,\quad p\cdot x = (u-u_0)\,,\quad p^ap_a= 0\,.
}{boost:8}
The quantity $X^+_0$ is a scale parameter essentially defining the units in which energy is being measured. It can be fixed to 1 in ``natural units''. The other directional derivative of the dilaton, $X^-$, vanishes in the UR limit because the conservation equation following from (\ref{eq:a5}), (\ref{eq:a6})
\eq{
X^{-a}X^+X^-=-M+\frac{B}{2(b+1)}X^{b+1}
}{boost:43}
implies that $X^-$ has to vanish for nonvanishing $X^+$ in the double limit $M,B\to 0$ (in fact, the limit $B\to 0$ is sufficient because $M$ scales with $\sqrt{B}$, cf.\ eq.\ (5.11) of \cite{Grumiller:2002nm}).

It is illuminating to discuss the patching procedure from a different perspective: suppose, for definiteness, that for $M>0$ a Killing horizon is present. Take an EF patch $X^+>0$ where the horizon is at $X^-=0$. If one performs a boost with $\ga\to-\infty$ then everywhere this transformation cancels, except at the horizon $u=u_0$ (simply because $X^-$ remains zero, as much as it would like to get boosted) where additionally $X^+$ vanishes in this limit. By continuity one can patch a whole region $u_1<u<u_0$ (a blow-up of the shock wave) with this feature which will be nothing but a constant dilaton vacuum (CDV), i.e.\ a solution with $X^+=0=X^-$, $X=\rm const.$,\footnote{\label{fn:4} The limit $B\to 0$ has to be employed to guarantee consistency with the equations of motion. This agrees with the scaling behaviour of the mass $M\to 0$ reatining a finite ADM energy (e.g.\ as given by eq.\ (5.11) of \cite{Grumiller:2002nm}; the quantity $-\mathcal{C}_0$ in that equation corresponds to $M$ in the present work).} while for $u>u_0$ the standard solution (\ref{boost:2}) can be applied in the limit $M\to 0$. Beyond $u<u_1$ the second EF patch $X^+<0$ can be pasted. To undo the blow-up one has to take the limit $u_1\to u_0$.

\subsection{The line element}

The line element is very simple, provided the background geometry is either flat (MGS models, $a=b+1$), Rindler ($b=0$) or (Anti)deSitter ($a+b=1$): the ensuing Killing norm is of the form $A_2r^2+A_1r+A_0$ and the curvature scalar is given by $2A_2$. Thus, for the cases under consideration the line element of a generic CDV (which can only be flat, Rindler, or (A)dS) can be patched to (\ref{boost:2}) in the limit $M\to 0$ such that curvature is continuous (it is a global constant for (A)dS and zero for Rindler or MGS). The metric is globally given by\footnote{More general cases -- neither MGS, nor Rindler, nor (A)dS -- will be addressed in sect.\ \ref{se:4}.}
\eq{
(ds)^2_{\rm boost} = 2dudr+\left(A_2r^2+A_1r+A_0\right)(du)^2\,,\quad A_i\in\mathbb{R}\,.
}{boost:9}
However, if this is claimed to be the nontrivial UR limit where is, indeed, the nontriviality? After all, if we take, say, spherically reduced Schwarzschild and perform the indicated steps only Minkowski spacetime is obtained, but there does not seem to be a trace of a shock wave. 

Consequently, it is somewhat puzzling that all curvature invariants are continuous but nevertheless a shock wave emerges as displayed by (\ref{boost:7}). This possibility is a special feature of dilaton gravity.\footnote{In Einstein gravity obviously continuity of the Ricci tensor implies continuity of the energy-momentum tensor. In dilaton gravity the dilaton field deforms the way in which gravity interacts with matter. This is displayed clearly in eqs.\ (\ref{eq:a6}), (\ref{eq:a7}) and (\ref{eq:a9}).} It has been observed previously \cite{Grumiller:2003ad} in the context of the kink solution of the dimensionally reduced gravitational Chern-Simons term \cite{Guralnik:2003we}. As noted before, {\em the energy-momentum 1-form is not induced by non-smoothness in curvature, but by a jump in one of the directional derivatives} $X^\pm$ {\em of the dilaton field} $X$ entering the equation of motion (\ref{eq:a6}). It is worthwhile mentioning that only (anti-)selfdual matter fluxes can be generated in this way. Although they do not affect the line-element they deform the dilaton profile (the dilaton as a function of the non-Killing coordinate) which aquires a kink. So if one thinks of the dilaton as being part of geometry (as it would be natural in the context of dimensionally reduced theories)\footnote{Even for generic dilaton theories (not neccessarily in $2D$) it is sometimes possible to interpret the dilaton geometrically in terms of a volume element density independent from the metric \cite{Graf:2002tw}.} then, indeed, geometry is deformed in the UR limit, albeit the intrinsically $2D$ line element is not.

\subsection{Geodesics of test particles}

As a direct consequence of the continuity of the curvature geodesics of test particles crossing the critical line $u=u_0$ do {\em not} receive a kick from the shock, i.e.\ they pass continuously and without kink through it. This is quite different from the behavior of test particles crossing the AS shock \cite{Balasin:1997mq}. The reason for this discrepancy is of course that the shock is not generated by the twodimensional metric but rather by the dilaton field. Therefore, similar features as in four dimensions can only be expected if test particles are coupled to the dilaton field (as would be natural in the framework of dimensionally reduced theories). Alternatively -- in the Schwarzschild case -- one can ``undo'' the dimensional reduction, i.e.\ study test particles in four dimensions by adding the angular part $-Xd\Om^2$ to the (flat) twodimensional line element. Since this calculation is quite instructive we will provide it explicitly:
The geodesic equation $(u\cdot\nabla)u^\mu=0=(u^nD_n)(u^mE_m^\mu)$ can be decomposed according to the standard 2-2 split into an angular part (indices from the middle of the alphabet) and into an intrinsic $2D$ part (indices from the beginning of the alphabet) -- see for instance appendix A of \cite{Grumiller:2002nm}. Quantities with tilde on top are intrinsic quantities with respect to the reduced theory or the sphere, respectively. A brief calculation yields:
\begin{align}
(\tilde{u}^b\tilde{D}_b) \tilde{u}^a &= \frac12 X(\tilde{E}^a\ln{X})\tilde{u}^i\tilde{u}_i \label{boost:66}\\
(\tilde{u}^j\tilde{D}_j)\tilde{u}^i &= -(\tilde{E}_a\ln{X})\tilde{u}^a\tilde{u}^i\label{boost:67}
\end{align}
The angular equation (\ref{boost:67}) implies that $\tilde{u}^i\tilde{u}_i X^2$ has to be constant along a $4D$ geodesic. Inserting this result into (\ref{boost:66}) yields an effective force,
\eq{
(\tilde{u}^b\tilde{D}_b)\tilde{u}^a \propto \frac{\tilde{E}^aX}{X^2}\,,
}{boost:68}
which is proportional to the first derivative of the dilaton field $X$. Thus, if the latter jumps\footnote{The quantities $E^\pm X$ are proportional to $X^\pm$ and therefore the force jumps by a finite amount.} at the shock front the geodesics of $4D$ testparticles are continuous and differentiable but not their velocities (which are just continuous), even though the intrinsic $2D$ geometry might be smooth.

\section{Quantization on and of UR geometries}\label{se:qu}

When discussing the quantization we will exclusively refer to a massless scalar field $\phi$ which, however, can be coupled nonminimally to the dilaton field:
\begin{equation}
L^{(m)} = \frac 12\int_{\mathcal{M}_2}\;F(X)\, d \phi  \wedge \ast d \phi = 
\int_{\mathcal{M}_2}\; F(X)\, \phi^+ \wedge \phi^-\; ,
\label{eq:a70}
\end{equation}
with $\phi^\pm:=d\phi\mp\ast d\phi$. In the (anti-)selfdual case $\phi^+$ ($\phi^-$) vanishes, respectively. If $F(X)=\rm const.$ we will call the scalar field minimally coupled, otherwise nonminimally coupled.

\begin{figure}
\centering
\epsfig{file=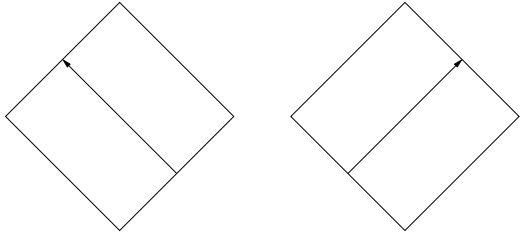,width=0.5\linewidth}
\caption{The two possible orientations of UR geometries}
\label{fig:2CP}
\end{figure}

First, we would like to present an intriguing reinterpretation of the UR limit: if one performs a path integration not over the scalar field $\phi$ but {\em separately} path integrates its (anti-)self dual components $\phi^\pm$ then one obtains, for instance, the anti-self duality condition $\phi^-=0$ as a constraint from the path integral over the self dual component. Of course, this path integration by no means is equivalent to the proper one over $\phi$. But it provides a somwhat unexpected tool to genereate the UR limit. In a sense, the components $\phi^\pm$ decouple from each other, much like fermions in the chiral limit do. Note that {\em either} $\phi^+$ {\em or} $\phi^-$ has to vanish as a constraint, depending on the order of integration, while the other one remains completely unconstrained. These two possible orders of path integration correspond to the two possible orientations of the Carter-Penrose diagram of the UR geometry, cf.\ fig.\ \ref{fig:2CP}.
It could be worthwhile to exploit this observation also in other contexts.

\subsection{Quantization of geometry}

Of course it is not possible to quantize geometry in a background independent way in the context of a specific background (the UR geometry). However, one can impose the limit $B\to 0$ discussed in footnote \ref{fn:4} already at the level of the action and try to quantize the resulting geometry in a background independent manner. In fact, this already has been done for generic dilaton gravity with matter \cite{Haider:1994cw,Grumiller:2002dm} 
and it remains to discuss the special features that arise due to $B=0$. Models of this type have been called ``generalized teleparallel theories'' in ref.\ \cite{Grumiller:2002dm} and they are particularly simple (e.g.\ the virtual black hole phenomenon is absent for these models). The reason for this simplicity (or rather triviality) is that assuming the UR limit already in the action before the quantization eliminates all interesting perturbations. Thus, it will be more relevant to discuss quantization of matter {\em on} such a background.

\subsection{Quantization on geometry}

For minimally coupled matter quantization on the UR background yields the same result as quantization on the ground state geometry because the kink in the dilaton profile is not felt by the modes of the scalar field (in the same way that intrinsically twodimensional test particles did not feel the shock wave).

However, for nonminimal coupling new effects can be expected. Indeed, the 1-loop effective action (6.42) of \cite{Grumiller:2002nm} 
\begin{equation}
\label{ehnWc}
W=\frac{1}{96\pi }\int d^{2}x\sqrt{-g}\left[ R\frac{1}{\Box }R-12(\nabla \Phi )^{2}\frac{1}{\Box }R+12R\Phi \right] 
\end{equation}
employed by many authors \cite{Bousso:1998wi} 
contains derivative terms of the dilaton und thus $\theta$-and $\de$-functions will arise. Therefore, we are going to discuss this issue in technical detail following the approach of Christensen and Fulling \cite{Christensen:1977jc} in full analogy to sect.\ 6.2 of \cite{Grumiller:2002nm} (cf.\ also references therein). In order to match with the definitions used in that section the redefinition $X=\exp{(-2\Phi)}$ will be applied. We will restrict to the MGS case; thus, there is no background curvature and any nontrivial effect will be solely due to the kink in the dilaton profile. 

The zeroth step is to check whether or not the classical trace of the energy-momentum tensor vanishes. Obviously this is the case for (\ref{boost:7}) as $\mom^+$ contains only a flux component $\mom^{++}$ but no trace component $\mom^{+-}$.

The first goal is to obtain the anomalous trace, e.g.\ by heat kernel methods and $\zeta$-function regularization (for a recent review cf.\ \cite{Vassilevich:2003xt}). Since all distributions enter only linearly the smeared $\zeta$-function has a well-defined distributional limit (heat kernel methods have been applied successfully in the presence of singular boundaries in the context of the brane-world scenario \cite{Gilkey:2001mj}; also there, as a rule of thumb, if an expression is well-defined, i.e.\ no ``squares of $\de$-functions'' arise, the heat kernel coefficients can be calculated). For vanishing background curvature the anomalous trace reads \cite{Mukhanov:1994ax} 
\eq{
\mom^\mu{}_\mu=2\mom_{z\bar{z}} = \frac{1}{4\pi}\left(\nabla^2\Phi-(\nabla\Phi)^2\right)=-\frac1\pi\left(\frac{\nabla^2X}{X}+(\nabla\ln{X})^2\right)\,.
}{boost:80}
Since the dilaton is only continuous it seems that the anomalous trace acquires a $\de$-function and possibly a $\theta^2$ contribution. To simplify the discussion lightcone gauge will be employed, $(ds)^2=2dzd\bar{z}$, $u=z$, $\nabla^2=2\partial_z\partial_{\bar{z}}$, and $\mom^+_z=\mom_{-z}=\mom_{--}=\mom_{zz}=\mom^{\bar{z}\bar{z}}$. Since the dilaton depends only on $z$ the first term in (\ref{boost:80}) disappears. The second term vanishes because it is proportional to $X^+X^-$ and $X^-=0$ in the UR limit. Thus, {\em there is no anomalous trace in the UR limit}.

The second step is to consider the energy-momentum (non-)conservation
\eq{
\nabla_\mu \mom^\mu{}_\nu = -(\partial_\nu\Phi) \frac{1}{\sqrt{-g}}\frac{\de W}{\de \Phi}
}{boost:81}
and to derive the flux component $\mom_{zz}$ from it. Obviously the term on the right hand side containing the 1-loop effective action (\ref{ehnWc}) vanishes for minimal coupling. If a flux exists it is usually called ``Hawking flux''. Its asymptotic limit is normally related to a Hawking temperature $T_H$ by virtue of the $2D$ Stefan-Boltzmann law $\mom_{zz}|_{\rm asy.}=T_H^2\pi/6$.

In the present case the left hand side of (\ref{boost:81}) simplifies to $\partial_{\bar{z}} T_{zz}$ while the right hand side contains an antiholomorphic integration constant $g(\bar{z})$ coming from $(1/\square) R$ (the holomorphic contribution to this term vanishes after differentiation) 
\eq{
\partial_{\bar{z}} \mom_{zz} = \frac{1}{32\pi}\frac{(X^+)^2}{X^2} g'(\bar{z})\,.
}{boost:82}
Thus, the flux component is given by
\eq{
\mom_{zz} = \frac{1}{32\pi}\frac{(X^+)^2}{X^2} g(\bar{z}) + f(z)\,.
}{boost:84}
The holomorphic function $f(z)$ can be fixed\footnote{``Normally'' at this point there is just a constant to be fixed the choice of which corresponds to the selection of a vacuum (e.g.~ Hartle-Hawking, Boulware or Unruh; cf.~e.g.~\cite{Kummer:1999zy}). Also in the present case the fixing of $f(z)$ can be considered as the selection of a certain vacuum. The only preferred choice seems to be the one given in the main text. We are grateful to D.V.~Vassilevich for discussions on this point.} such that it coincides with the classical flux in (\ref{boost:7}). 

Since the background geometry -- {\em including the dilaton} -- allows for a Killing vector $\partial_{\bar{z}}$ this symmetry should be obeyed by the energy-momentum tensor. Thus, the antiholomorphic part of $(1/\square) R$ has to vanish and therefore $g(\bar{z})=0$. As a result, the only flux at 1-loop level is given by the {\em classical} flux. In a sense, the shock wave emerging from/as the UR BH provides its own Hawking radiation.

\section{Discussion}\label{se:4}

Exploiting the double cover structure of Carter-Penrose diagrams, the UR limit of the $ab$-family (\ref{cs:1.5}) of twodimensional dilaton gravity (\ref{cs:1}) has been constructed. Although the line element (\ref{boost:9}) is twice differentiable for Minkowski-, Rindler-, or (A)dS-ground state theories (i.e.\ models where the ground state geometry (\ref{boost:3}) behaves correspondingly) and thus curvature is continuous the energy-momentum 1-form acquires a distributional term (\ref{boost:7}) resembling the analogous expression in higher dimensions. Alternative approaches to construct the UR limit which work well in $D=4$ did not seem to work straightforwardly for generic dilaton gravity.

Clearly, the proposed procedure works also for more generic dilaton gravity models, provided the related Carter-Penrose diagramm exhibits the double covering feature. For convenience, here is again the prescription:
\blist
\item Take two copies of patches of the ground state solution (vanishing Casimir) of (\ref{eq:FOG}) for a given $\mathcal V$, one with $X^+>0$ and one with $X^+<0$ in the whole patch ($X^+$ has been chosen for definiteness -- of course, the same procedure works as well for $X^-$; this freedom corresponds to the two possible orientations of the boosted geometry displayed in fig.\ \ref{fig:2CP})
\item Glue them together along a $u=\rm const.$ line, i.e.\ solve eq.\ (\ref{eq:a6}) with (\ref{boost:6}) and extract the ensuing energy-momentum 1-form; it will always be equivalent to the one in (\ref{boost:7}), regardless of the underlying model; the factor $X^+_0\neq 0$ defines the energy scale
\item The line element is globally given by the one for the ground state solution with the possible exception of the light like line $u=u_0$ where the shock wave is located
\elist
There is a slight caveat for geometries which are not of the form (\ref{boost:9}): although the patching works in the same way the blow up scenario discussed below eq.\ (\ref{boost:8}) does not work anymore. However, one can adopt the viewpoint that nevertheless the ``correct'' UR limit is defined to be the patched geometry.

We leave it as an open problem how to define the UR limit for more complicated geometries (e.g.\ the solution G11 on p.\ 26 of the second ref.\ \cite{Klosch:1996fi}) that arise in the context of twodimensional dilaton gravity.

We hope that our investigation of ultrarelativistic shock-waves clarified some of the similarities as well as the differences between the $4D$ and the $2D$ situation.

\section*{Acknowledgement}
 
This work has been supported by projects P-14650-TPH and J2330-N08 of the Austrian Science Foundation (FWF). DG is grateful to W.~Kummer and D.V.~Vassilevich for a fruitful long-time collaboration on dilaton gravity. Part of this project was performed during a workshop at the International Erwin-Schr\"odinger Institute in Vienna.

\begin{appendix}

\section{Equations of motion}\label{app:A}

The equations of motion derived from (\ref{cs:1}) read
\begin{eqnarray}
 &  & dX+X^{-}e^{+}-X^{+}e^{-}=0\, ,\label{eq:a5} \\
 &  & (d\pm \omega )X^{\pm }\pm \left(\frac{a X^+X^-}{X}+\frac{B}{2}X^{a+b}\right)e^{\pm }+\mom^{\pm }=0\, ,\label{eq:a6} \\
 &  & d\omega +\epsilon \left(\frac{a X^+X^-}{X^2}-\frac{B(a+b)}{2}X^{a+b-1}\right)+\mom=0\, ,\label{eq:a7} \\
 &  & (d\pm \omega )e^{\pm }-\epsilon \, \frac{a X^\pm}{X}=0\, .\label{eq:a8} 
\end{eqnarray}
The quantities \eq{
\mom^{\pm }:= \delta L^{(m)}/\delta e^{\mp}, \quad \mom := \delta L^{(m)}/\delta X
}{eq:a9} 
contain the couplings to matter where $L^{(m)}$ is the (unspecified) matter Lagrangian 2-form. $\mom^\pm$ is the energy-momentum 1-form.

It is useful to study the continuity properties: if the curvature scalar is continuous then connection and torsion must be differentiable and the dyad must be twice differentiable (of course the metric inherits this property). If $X^+$ and/or $X^-$ is discontinuous then the dilaton profile has a kink, but it remains continuous; the energy-momentum 1-form $\mom^\pm$ acquires a $\de$-function.

Note that the first equation (\ref{eq:a5}) allows to interpret the Lagrange multipliers $X^\pm$ as directional derivatives of the dilaton by contracting it with $E_\pm$ (using $dX=e^a E_a X$): $X^\pm = \pm E_\mp X$.

\section{Distributional curvature}\label{app:B}

For simplicity we restrict ourselves to MGS models. Then, the only source for a singularity is the factor $aX^+X^-/X$ in the action (\ref{cs:1}). It can be regularized, for instance, in the following manner:
\eq{
U_\eps(X) := -\frac{aX}{X^2+\eps^2}\,,\quad \lim_{\eps\to 0}U_\eps(X)=U(X)=-\frac{a}{X}
}{boost:10}
The regularized theory is still a dilaton gravity model, albeit not of the $a-b$ family anymore. The minus sign in (\ref{boost:10}) has been adjusted for consistency with the notation in \cite{Grumiller:2002nm}. The corresponding curvature scalar\footnote{We calculate the curvature scalar related to the Levi-Civit\'a connection, but not the one related to $\om$ appearing in the equations of motion.} turns out as
\eq{
R_\eps(X)= -aM(X^2-\eps^2)(X^2+\eps^2)^{a/2-2}\,.
}{boost:11} 
The dilaton and the radial coordinate are now related by
\eq{
dr=(X^2+\eps^2)^{-a/2}dX\,.
}{boost:13}
For finite $\eps$ this provides a regular distribution proportional to the mass $M$
\eq{
(R_\eps,\phi):=\int\limits_{-\infty}^{\infty}dr\phi(r)R_\eps(X(r))=-aM\int\limits_{-\infty}^{\infty}dX\phi(X)\frac{X^2-\eps^2}{(X^2+\eps^2)^2}\,.
}{boost:12}
Some remarks are in order: first of all, the range of integration could also be argued to be $(0,\infty)$; second, one might be tempted to take the dilaton $dX$ as integration variable and not the radial coordinate $dr$ -- however, only with respect to the latter is the line element well defined in the asymptotic region; third, the change from an integration over $r$ to an integration over $X$ in (\ref{boost:12}) might be questioned on the grounds that integration over a ``radius'' leads to a different behaviour than integration over the ``surface area''; finally, one could in principle introduce by hand a dilaton dependent factor, quasi as part of the measure -- this is motivated by the fact that spherical reduction from $D=4$ to $D=2$ implies $\sqrt{-g^{(4)}}=|X|\sqrt{-g^{(2)}}$. If, despite of these remarks, one proceeds with (\ref{boost:12}) first of all some subtraction and rescaling has to be performed ($X=y\eps$):
\meq{
(R_\eps,\phi):=-aM\int\limits_{-\infty}^{\infty}dX\Big(\phi(X)-\theta(\mu-|X|)(\phi(0)+X\phi'(0))\Big)\frac{X^2-\eps^2}{(X^2+\eps^2)^2}\\
+\frac{aM}{\eps}\int\limits_{-\mu/\eps}^{\mu/\eps} dy(\phi(0)+y\eps\phi'(0))\frac{y^2-1}{(y^2+1)^2}\,.
}{boost:14}
Both integrals allow for a well-defined limit $\eps\to 0$:
\eq{
\lim_{\eps\to 0}(R_\eps,\phi)=-aM\left(\left[\frac{1}{X^2}\right]_\mu,\phi\right)-\frac{2aM}{\mu} \left(\de(X),\phi\right)
}{boost:15}
The finite ambiguity encoded in $\mu$ can be treated as follows:
\meq{
\left(\left[\frac{1}{X^2}\right]_\mu,\phi\right)=\left(\left[\frac{1}{X^2}\right]_1,\phi\right)+\int\limits_{-\infty}^{\infty}dX\Big(\phi(0)+X\phi'(0)\Big)\\
\Big(\theta(1-|X|)-\theta(\mu-|X|)\Big)\frac{1}{X^2} = \left(\left[\frac{1}{X^2}\right]_1,\phi\right) + 2 \left(1-\frac1\mu\right)\left(\de(X),\phi\right)
}{boost:16} 
Thus, the $\mu$-dependence in the whole term simply cancels and the final result for the curvature scalar of MGS theories as regular distribution is given by
\eq{
(R,\phi) = -aM\left(\left(\left[\frac{1}{X^2}\right]_1,\phi\right)+2\left(\de(X),\phi\right)\right)\,. 
}{boost:17}

However, this result cannot be used directly to obtain information about the energy-momentum 1-form $\mom^\pm$ because in dilaton gravity only the quantity $\mom$ as defined in (\ref{eq:a9}) couples to curvature. This is not the case in the fourdimensional version where one can successfully apply distributional methods to derive the energy-momentum tensor of the Kerr-Newman spacetime family \cite{Balasin:1994kf} the knowledge of which allows for an unambiguous UR limit, e.g.\ the AS metric (\ref{eq:AS}) in the Schwarzschild case \cite{Balasin:1995tb}.

Thus, the direct method advocated in the main text seems to be superior in the context of $2D$ dilaton gravity.

\end{appendix}


\begin{thebibliography}{10}

\bibitem{Aichelburg:1971dh}
P.~C. Aichelburg and R.~U. Sexl, ``On the gravitational field of a massless
  particle,'' {\em Gen. Rel. Grav.} {\bf 2} (1971)
303--312.

\bibitem{Jordan:1960}
P.~Jordan, J.~Ehlers, and W.~Kundt, ``{Strenge L{\"o}sungen der Feldgleichungen
  der Allgemeinen Relativit{\"a}tstheorie},'' {\em Akad. Wiss. Lit. (Mainz)
  Abhandl. Math.-Nat. Kl.} {\bf 2} (1960) 21.

\bibitem{Dray:1984ha}
T.~Dray and G.~'t~Hooft, ``The gravitational shock wave of a massless
  particle,'' {\em Nucl. Phys.} {\bf B253} (1985)
173.

\bibitem{Lousto:1988ua}
C.~O. Lousto and N.~Sanchez, ``The curved shock wave space-time of
  ultrarelativistic charged particles and their scattering,'' {\em Int. J. Mod.
  Phys.} {\bf A5} (1990)
915;
``The ultrarelativistic limit of the
  {K}err-{N}ewman geometry and particle scattering at the {P}lanck scale,''
  {\em Phys. Lett.} {\bf B232} (1989)
462.

\bibitem{Balasin:1995tb}
H.~Balasin and H.~Nachbagauer, ``{The Ultrarelativistic Kerr geometry and its
  energy momentum tensor},'' {\em Class. Quant. Grav.} {\bf 12} (1995)
  707--714,
\href{http://www.arXiv.org/abs/gr-qc/9405053}{{\tt gr-qc/9405053}}.

\bibitem{Balasin:1995tj}
H.~Balasin and H.~Nachbagauer, ``{Boosting the Kerr geometry into an arbitrary
  direction},'' {\em Class. Quant. Grav.} {\bf 13} (1996) 731--738,
\href{http://www.arXiv.org/abs/gr-qc/9508044}{{\tt gr-qc/9508044}};
C.~Barrabes and P.~A. Hogan, ``{Light-Like Boost of the Kerr Gravitational
  Field},'' {\em Phys. Rev.} {\bf D67} (2003) 084028,
\href{http://www.arXiv.org/abs/gr-qc/0303055}{{\tt gr-qc/0303055}}.

\bibitem{Alonso:1987}
R.~Alonso and N.~Zamorano, ``{Generalized Kerr-Schild metric for a massless
  particle on the Reissner-Norstr{\"o}m horizon},'' {\em Phys.Rev.} {\bf D35}
  (1987) 1798;
K.~Sfetsos, ``On gravitational shock waves in curved space-times,'' {\em Nucl.
  Phys.} {\bf B436} (1995) 721--746,
\href{http://www.arXiv.org/abs/hep-th/9408169}{{\tt hep-th/9408169}}.

\bibitem{Balasin:1999wg}
H.~Balasin, ``{Generalized Kerr-Schild metrics and the gravitational field of a
  massless particle on the horizon},'' {\em Class. Quant. Grav.} {\bf 17}
  (2000) 1913--1920,
\href{http://www.arXiv.org/abs/gr-qc/9909082}{{\tt gr-qc/9909082}}.

\bibitem{Mandal:1991tz}
G.~Mandal, A.~M. Sengupta, and S.~R. Wadia, ``Classical solutions of
  two-dimensional string theory,'' {\em Mod. Phys. Lett.} {\bf A6} (1991)
1685--1692;
S.~Elitzur, A.~Forge, and E.~Rabinovici, ``Some global aspects of string
  compactifications,'' {\em Nucl. Phys.} {\bf B359} (1991)
581--610;
E.~Witten, ``On string theory and black holes,'' {\em Phys. Rev.} {\bf D44}
  (1991)
314--324;
R.~Dijkgraaf, H.~Verlinde, and E.~Verlinde, ``String propagation in a black
  hole geometry,'' {\em Nucl. Phys.} {\bf B371} (1992)
269--314.

\bibitem{Callan:1992rs}
C.~G. Callan, Jr., S.~B. Giddings, J.~A. Harvey, and A.~Strominger,
  ``Evanescent black holes,'' {\em Phys. Rev.} {\bf D45} (1992) 1005--1009,
\href{http://www.arXiv.org/abs/hep-th/9111056}{{\tt hep-th/9111056}}.

\bibitem{Barbashov:1979}
B.~M. Barbashov, V.~V. Nesterenko, and A.~M. Chervyakov, ``The solitons in some
  geometrical field theories,'' {\em Teor. Mat. Fiz.} {\bf 40} (1979)
15--27;
{\em J. Phys.} {\bf A13} (1979)
301--312;
{\em Theor. Math. Phys.} {\bf 40} (1979)
572--581;
E.~D'Hoker and R.~Jackiw, ``Liouville field theory,'' {\em Phys. Rev.} {\bf
  D26} (1982)
3517;
C.~Teitelboim, ``Gravitation and {H}amiltonian structure in two space-time
  dimensions,'' {\em Phys. Lett.} {\bf B126} (1983)
41;
E.~D'Hoker, D.~Freedman, and R.~Jackiw, ``{SO}(2,1) invariant quantization of
  the {L}iouville theory,'' {\em Phys. Rev.} {\bf D28} (1983)
2583;
E.~D'Hoker and R.~Jackiw, ``Space translation breaking and compactification in
  the {L}iouville theory,'' {\em Phys. Rev. Lett.} {\bf 50} (1983)
1719--1722;
R.~Jackiw, ``Another view on massless matter - gravity fields in two-
  dimensions,''
\href{http://arXiv.org/abs/hep-th/9501016}{{\tt hep-th/9501016}}.


\bibitem{Douglas:2003up}
N.~Berkovits, S.~Gukov, and B.~C. Vallilo, ``{Superstrings in 2D backgrounds
  with R-R flux and new extremal black holes},'' {\em Nucl. Phys.} {\bf B614}
  (2001) 195--232,
\href{http://arXiv.org/abs/hep-th/0107140}{{\tt hep-th/0107140}};
M.~R. Douglas {\em et al.}, ``A new hat for the c = 1 matrix model,''
\href{http://www.arXiv.org/abs/hep-th/0307195}{{\tt hep-th/0307195}};
D.~M. Thompson, ``{AdS solutions of 2D type 0A},''
\href{http://www.arXiv.org/abs/hep-th/0312156}{{\tt hep-th/0312156}};
S.~Gukov, T.~Takayanagi, and N.~Toumbas, ``Flux backgrounds in 2d string
  theory,''
\href{http://www.arXiv.org/abs/hep-th/0312208}{{\tt hep-th/0312208}};
J.~Davis, L.~A.~P. Zayas, and D.~Vaman, ``On black hole thermodynamics of 2-d
  type {0A},'' {\em JHEP} {\bf 03} (2004) 007,  
\href{http://www.arXiv.org/abs/hep-th/0402152}{{\tt hep-th/0402152}};
U.~H. Danielsson, J.~P. Gregory, M.~E. Olsson, P.~Rajan, and M.~Vonk, ``Type {0A}
  2d black hole thermodynamics and the deformed matrix model,''
\href{http://www.arXiv.org/abs/hep-th/0402192}{{\tt hep-th/0402192}}.

\bibitem{Grumiller:2002nm}
D.~Grumiller, W.~Kummer, and D.~V. Vassilevich, ``Dilaton gravity in two
  dimensions,'' {\em Phys. Rept.} {\bf 369} (2002) 327--429,
\href{http://arXiv.org/abs/hep-th/0204253}{{\tt hep-th/0204253}}.

\bibitem{Schaller:1994es}
P.~Schaller and T.~Strobl, ``Poisson structure induced (topological) field
  theories,'' {\em Mod. Phys. Lett.} {\bf A9} (1994) 3129--3136,
\href{http://arXiv.org/abs/hep-th/9405110}{{\tt hep-th/9405110}}.

\bibitem{Katanaev:1997ni}
M.~O. Katanaev, W.~Kummer, and H.~Liebl, ``On the completeness of the black
  hole singularity in 2d dilaton theories,'' {\em Nucl. Phys.} {\bf B486}
  (1997) 353--370,
\href{http://www.arXiv.org/abs/gr-qc/9602040}{{\tt gr-qc/9602040}}.

\bibitem{Klosch:1996fi}
T.~Kl{\"o}sch and T.~Strobl, ``Classical and quantum gravity in
  (1+1)-dimensions. {P}art {I}: {A} unifying approach,'' {\em Class. Quant.
  Grav.} {\bf 13} (1996) 965--984,
\href{http://www.arXiv.org/abs/arXiv:gr-qc/9508020}{{\tt arXiv:gr-qc/9508020}};
``Classical and quantum gravity in 1+1 dimensions.
  {P}art {II}: {T}he universal coverings,'' {\em Class. Quant. Grav.} {\bf 13}
  (1996) 2395--2422,
\href{http://www.arXiv.org/abs/arXiv:gr-qc/9511081}{{\tt arXiv:gr-qc/9511081}};
``{Classical and quantum gravity in 1+1
  dimensions. Part III: Solutions of arbitrary topology and kinks in 1+1
  gravity},'' {\em Class. Quant. Grav.} {\bf 14} (1997) 1689--1723,
\href{http://arXiv.org/abs/hep-th/9607226}{{\tt hep-th/9607226}};
``A global view of kinks in 1+1 gravity,'' {\em
  Phys. Rev.} {\bf D57} (1998) 1034--1044,
\href{http://www.arXiv.org/abs/arXiv:gr-qc/9707053}{{\tt arXiv:gr-qc/9707053}}.

\bibitem{Grumiller:2003ad}
D.~Grumiller and W.~Kummer, ``{The classical solutions of the dimensionally
  reduced gravitational Chern-Simons theory},'' {\em Ann. Phys.} {\bf 308}
  (2003) 211--221,
\href{http://www.arXiv.org/abs/hep-th/0306036}{{\tt hep-th/0306036}}.

\bibitem{Guralnik:2003we}
G.~Guralnik, A.~Iorio, R.~Jackiw, and S.~Y. Pi, ``{Dimensionally reduced
  gravitational Chern-Simons term and its kink},'' {\em Ann. Phys.} {\bf 308} (2003) 222--236,
\href{http://www.arXiv.org/abs/hep-th/0305117}{{\tt hep-th/0305117}}.

\bibitem{Graf:2002tw}
W.~Graf, ``Geometric dilaton gravity and smooth charged wormholes,'' {\em Phys.
  Rev.} {\bf D67} (2003) 024002,
\href{http://www.arXiv.org/abs/gr-qc/0209002}{{\tt gr-qc/0209002}}.

\bibitem{Balasin:1997mq}
H.~Balasin, ``Geodesics for impulsive gravitational waves and the
  multiplication of distributions,'' {\em Class. Quant. Grav.} {\bf 14} (1997)
  455--462,
\href{http://www.arXiv.org/abs/gr-qc/9607076}{{\tt gr-qc/9607076}}.

\bibitem{Haider:1994cw}
F.~Haider and W.~Kummer, ``{Quantum functional integration of nonEinsteinian
  gravity in d = 2},'' {\em Int. J. Mod. Phys.} {\bf A9} (1994)
207--220;
W.~Kummer, H.~Liebl, and D.~V. Vassilevich, ``Exact path integral quantization
  of generic 2-d dilaton gravity,'' {\em Nucl. Phys.} {\bf B493} (1997)
  491--502,
\href{http://arXiv.org/abs/gr-qc/9612012}{{\tt gr-qc/9612012}};
``Non-perturbative path integral of
  2d dilaton gravity and two-loop effects from scalar matter,'' {\em Nucl.
  Phys.} {\bf B513} (1998) 723--734,
\href{http://arXiv.org/abs/hep-th/9707115}{{\tt hep-th/9707115}};
``Integrating geometry in general
  2d dilaton gravity with matter,'' {\em Nucl. Phys.} {\bf B544} (1999)
  403--431,
\href{http://www.arXiv.org/abs/hep-th/9809168}{{\tt hep-th/9809168}};
D.~Grumiller, W.~Kummer, and D.~V. Vassilevich, ``The virtual black hole in 2d
  quantum gravity,'' {\em Nucl. Phys.} {\bf B580} (2000) 438--456,
\href{http://www.arXiv.org/abs/gr-qc/0001038}{{\tt gr-qc/0001038}};
P.~Fischer, D.~Grumiller, W.~Kummer, and D.~V. Vassilevich, ``S-matrix for
  s-wave gravitational scattering,'' {\em Phys. Lett.} {\bf B521} (2001)
  357--363, \href{http://arXiv.org/abs/gr-qc/0105034}{{\tt gr-qc/0105034}};
Erratum ibid. {\bf B532} (2002) 373;
D.~Grumiller, {\em Quantum dilaton gravity in two dimensions with matter}.
\newblock PhD thesis, {T}echnische {U}niversit{\"a}t {W}ien, 2001.
\newblock
\href{http://www.arXiv.org/abs/gr-qc/0105078}{{\tt gr-qc/0105078}};
\newblock
``Virtual black hole phenomenology from 2d dilaton theories,''
  {\em Class. Quant. Grav.} {\bf 19} (2002) 997--1009,
\href{http://arXiv.org/abs/gr-qc/0111097}{{\tt gr-qc/0111097}}.

\bibitem{Grumiller:2002dm}
D.~Grumiller, W.~Kummer, and D.~V. Vassilevich, ``Virtual black holes in
  generalized dilaton theories (and their special role in string gravity),''
  {\em European Phys. J.} {\bf C30} (2003) 135--143,
\href{http://arXiv.org/abs/hep-th/0208052}{{\tt hep-th/0208052}}.

\bibitem{Bousso:1998wi}
R.~Bousso and S.~W. Hawking, ``{(Anti-)evaporation of Schwarzschild-de Sitter
  black holes},'' {\em Phys. Rev.} {\bf D57} (1998) 2436--2442,
\href{http://arXiv.org/abs/hep-th/9709224}{{\tt hep-th/9709224}};
R.~Bousso, ``Proliferation of de {S}itter space,'' {\em Phys. Rev.} {\bf D58}
  (1998) 083511,
\href{http://arXiv.org/abs/hep-th/9805081}{{\tt hep-th/9805081}};
``Quantum global structure of de {S}itter space,'' {\em Phys. Rev.}
  {\bf D60} (1999) 063503,
\href{http://arXiv.org/abs/hep-th/9902183}{{\tt hep-th/9902183}};
M.~Buric and V.~Radovanovic, ``{ADM mass of the quantum corrected Schwarzschild
  black hole},'' {\em Class. Quant. Grav.} {\bf 17} (2000) 33--42,
\href{http://arXiv.org/abs/gr-qc/9907106}{{\tt gr-qc/9907106}};
``{Quantum corrections for the
  Reissner-Nordstr{\"o}m black hole},'' {\em Class. Quant. Grav.} {\bf 16}
  (1999) 3937--3951,
\href{http://arXiv.org/abs/gr-qc/9907036}{{\tt gr-qc/9907036}};
A.~J.~M. Medved and G.~Kunstatter, ``Quantum corrections to the thermodynamics
  of charged 2-d black holes,'' {\em Phys. Rev.} {\bf D60} (1999) 104029,
\href{http://arXiv.org/abs/hep-th/9904070}{{\tt hep-th/9904070}};
S.~Nojiri and S.~D. Odintsov, ``Quantum dilatonic gravity in d = 2, 4 and 5
  dimensions,'' {\em Int. J. Mod. Phys.} {\bf A16} (2001) 1015--1108,
\href{http://arXiv.org/abs/hep-th/0009202}{{\tt hep-th/0009202}};
more references can be found in sect.\ 6 of ref.\ \cite{Grumiller:2002nm}.

\bibitem{Christensen:1977jc}
S.~M. Christensen and S.~A. Fulling, ``Trace anomalies and the {H}awking
  effect,'' {\em Phys. Rev.} {\bf D15} (1977)
2088--2104.

\bibitem{Vassilevich:2003xt}
D.~V. Vassilevich, ``Heat kernel expansion: User's manual,'' {\em Phys. Rept.}
  {\bf 388} (2003) 279--360,
\href{http://www.arXiv.org/abs/hep-th/0306138}{{\tt hep-th/0306138}}.

\bibitem{Gilkey:2001mj}
P.~B. Gilkey, K.~Kirsten, and D.~V. Vassilevich, ``Heat trace asymptotics with
  transmittal boundary conditions and quantum brane-world scenario,'' {\em
  Nucl. Phys.} {\bf B601} (2001) 125--148,
\href{http://www.arXiv.org/abs/hep-th/0101105}{{\tt hep-th/0101105}}.

\bibitem{Mukhanov:1994ax}
V.~Mukhanov, A.~Wipf, and A.~Zelnikov, ``{On 4-D Hawking radiation from
  effective action},'' {\em Phys. Lett.} {\bf B332} (1994) 283--291,
\href{http://arXiv.org/abs/hep-th/9403018}{{\tt hep-th/9403018}};
T.~Chiba and M.~Siino, ``Disappearance of black hole criticality in
  semiclassical general relativity,'' {\em Mod. Phys. Lett.} {\bf A12} (1997)
709--718;
S.~Ichinose, ``{Weyl anomaly of 2D dilaton-scalar gravity and hermiticity of
  system operator},'' {\em Phys. Rev.} {\bf D57} (1998) 6224--6229,
\href{http://arXiv.org/abs/hep-th/9707025}{{\tt hep-th/9707025}};
W.~Kummer, H.~Liebl, and D.~V. Vassilevich, ``Hawking radiation for
  non-minimally coupled matter from generalized 2d black hole models,'' {\em
  Mod. Phys. Lett.} {\bf A12} (1997) 2683--2690,
\href{http://arXiv.org/abs/hep-th/9707041}{{\tt hep-th/9707041}};
J.~S. Dowker, ``Conformal anomaly in 2d dilaton-scalar theory,'' {\em Class.
  Quant. Grav.} {\bf 15} (1998) 1881--1884,
\href{http://arXiv.org/abs/hep-th/9802029}{{\tt hep-th/9802029}}.

\bibitem{Kummer:1999zy}
W.~Kummer and D.~V. Vassilevich, ``{Hawking radiation from dilaton gravity in
  (1+1) dimensions: A pedagogical review},'' {\em Annalen Phys.} {\bf 8} (1999)
  801--827,
\href{http://arXiv.org/abs/gr-qc/9907041}{{\tt gr-qc/9907041}}.

\bibitem{Balasin:1994kf}
H.~Balasin and H.~Nachbagauer, ``{Distributional energy momentum tensor of the
  Kerr-Newman space-time family},'' {\em Class. Quant. Grav.} {\bf 11} (1994)
  1453--1462,
\href{http://arXiv.org/abs/gr-qc/9312028}{{\tt gr-qc/9312028}}.

\end{thebibliography}

\providecommand{\href}[2]{#2}\begingroup\raggedright\endgroup

\end{document}